\documentclass[usenatbib,useAMS,preprint,psfig2]{mn2e}
\usepackage{times}
\usepackage{amsmath}
\usepackage{amssymb,amsfonts}
\usepackage{graphicx}
\usepackage{epstopdf}
\usepackage{verbatim}
\input psfig.sty

\def\gapprox{\lower.4ex\hbox{$\;\buildrel >\over{\scriptstyle\sim}\;$}}
\def\lapprox{\lower.4ex\hbox{$\;\buildrel <\over{\scriptstyle\sim}\;$}} \def\be{\begin{equation}}
\def\be{\begin{equation}}
\def\ee{\end{equation}}
\def\bea{\begin{eqnarray}}
\def\eea{\end{eqnarray}}
\catcode`\@=11 \font\tenmib=cmmib10 \font\tensyb=cmbsy10
\font\tenbi=cmmib10
\newfam\bifam 
\textfont\bifam=\tenbi  

\def\unboldmath{\everymath{}\everydisplay{}
          \textfont\@ne\teni
          \textfont\tw@\tensy
          }
\def\boldmath{$\!\!$\relax\everymath{\mit}\everydisplay{\mit}
        \textfont\@ne\tenmib
        \textfont\tw@\tensyb
        \relax}%
\catcode`\@=12

\begin{document}

\tolerance=10000

\def\lesssim{\mathrel{\hbox{\rlap{\hbox{\lower4pt\hbox{$\sim$}}}\hbox{$<$}}}}
\def\gtrless{\mathrel{\hbox{\rlap{\hbox{\lower3pt\hbox{$<$}}}\hbox{$>$}}}}

\def\ms{\ifmmode {\rm M_{\odot}} \else ${\rm M_{\odot}}$\fi}    
\def\na{\ifmmode \nu_{A} \else $\nu_{A}$\fi}    
\def\nk{\ifmmode \nu_{K} \else $\nu_{K}$\fi}    
\def\ns{\ifmmode \nu_{{\rm s}} \else $\nu_{{\rm s}}$\fi}
\def\no{\ifmmode \nu_{1} \else $\nu_{1}$\fi}    
\def\nt{\ifmmode \nu_{2} \else $\nu_{2}$\fi}    
\def\ntk{\ifmmode \nu_{2k} \else $\nu_{2k}$\fi}    
\def\dnmax{\ifmmode \Delta \nu_{max} \else $\Delta \nu_{2max}$\fi}
\def\ntmax{\ifmmode \nu_{2max} \else $\nu_{2max}$\fi}    
\def\nomax{\ifmmode \nu_{1max} \else $\nu_{1max}$\fi}    
\def\nn{\ifmmode \nu_{\rm NBO} \else $\nu_{\rm NBO}$\fi}    
\def\nh{\ifmmode \nu_{\rm HBO} \else $\nu_{\rm HBO}$\fi}    
\def\nqpo{\ifmmode \nu_{QPO} \else $\nu_{QPO}$\fi}    
\def\nz{\ifmmode \nu_{o} \else $\nu_{o}$\fi}    
\def\nht{\ifmmode \nu_{H2} \else $\nu_{H2}$\fi}    
\def\ns{\ifmmode \nu_{\rm s} \else $\nu_{\rm s}$\fi}    
\def\nb{\ifmmode \nu_{{\rm burst}} \else $\nu_{{\rm burst}}$\fi}
\def\nkm{\ifmmode \nu_{km} \else $\nu_{km}$\fi}    
\def\ka{\ifmmode \kappa \else \kappa\fi}    
\def\dn{\ifmmode \Delta\nu \else \Delta\nu\fi}    

\def\vk{\ifmmode v_{k} \else $v_{k}$\fi}    
\def\va{\ifmmode v_{A} \else $v_{A}$\fi}    
\def\vf{\ifmmode v_{ff} \else $v_{ff}$\fi}    

\def\rs{\ifmmode R_{s} \else $R_{s}$\fi}    
\def\ra{\ifmmode R_{A} \else $R_{A}$\fi}    
\def\rso{\ifmmode R_{S1} \else $R_{S1}$\fi}    
\def\rst{\ifmmode R_{S2} \else $R_{S2}$\fi}    
\def\rmm{\ifmmode R_{M} \else $R_{M}$\fi}    
\def\rco{\ifmmode R_{co} \else $R_{co}$\fi}    
\def\ris{\ifmmode {\rm R}_{{\rm ISCO}} \else $ {\rm R}_{{\rm ISCO}} $\fi}
\def\rsix{\ifmmode R_{6} \else $R_{6}$\fi}    

\def\object{ \rm }

\def\be{\begin{equation}}
\def\ee{\end{equation}}
\def\bea{\begin{eqnarray}}
\def\eea{\end{eqnarray}}
\def\c{\cite}

\def\et{ {\it et al. \; }}
\def\lan{ \langle}
\def\ran{ \rangle}
\def\ov{ \over}
\def\ep{ \epsilon}

\def\mdot{\ifmmode \dot M \else $\dot M$\fi}    
\def\mxd{\ifmmode \dot {M}_{x} \else $\dot {M}_{x}$\fi}
\def\med{\ifmmode \dot {M}_{Edd} \else $\dot {M}_{Edd}$\fi}
\def\bff{\ifmmode B_{f} \else $B_{f}$\fi}

\def\apj{\ifmmode ApJ \else ApJ\fi}    
\def\apjl{\ifmmode  ApJ \else ApJ\fi}    %
\def\aap{\ifmmode A\&A \else A\&A\fi}    %
\def\mnras{\ifmmode MNRAS \else MNRAS\fi}    %
\def\nat{\ifmmode Nature \else Nature\fi}
\def\prl{\ifmmode Phys. Rev. Lett. \else Phys. Rev. Lett.\fi}
\def\prd{\ifmmode Phys. Rev. D. \else Phys. Rev. D.\fi}

\def\sax{\ifmmode SAX J1808.4-3658 \else SAX J1808.4-3658 \fi}

\def\et{{\it et al.}}

\title[The  correlations between the twin  kHz QPO
 frequencies]{The  correlations between the twin  kHz QPO
 frequencies  of LMXBs}

 \author[C.M. Zhang et al.]{C.M. Zhang$^{1}$, H.X.  Yin$^{1}$, Y.H. Zhao$^{1}$,   F. Zhang$^{2}$, L.M.
   Song$^{2}$\\
 1. National Astronomical Observatories,
  Chinese Academy of Sciences, Beijing 100012, China, zhangcm@bao.ac.cn \\
   2. Astronomical Institute, Institute of  High Energy Physics, Chinese Academy of Sciences, Beijing,
            China}

\date{\today}

\maketitle

\begin{abstract}

 We analyzed the recently published   kHz QPO data
  in the neutron star low-mass X-ray binaries (LMXBs), in order to
  investigate the different correlations of the twin peak kilohertz quasi-periodic
  oscillations (kHz QPOs) in bright Z sources and in the less luminous Atoll sources.
  We find that a power-law relation $\no\sim\nt^{b}$ between
  the upper  and the lower kHz QPOs with different indices:
  $b\simeq$1.5 for the Atoll source 4U 1728-34 and
  $b\simeq$1.9 for the Z source Sco X-1.
 The implications of our results for the theoretical models for kHz QPOs
 are discussed.

 \end{abstract}


 \date{Received~~2005 month day; accepted~~2005~~month day}

\section{INTRODUCTION}

With the advent of the Rossi X-ray Timing Explorer ({\em RXTE}),
our knowledge of the properties of the aperiodic variability of
 neutron star (NS) low mass  X-ray binaries (LMXBs)
 took a substantial step forward, especially initiated by the
 discovery of the  kilohertz quasi-periodic oscillations (kHz
 QPOs) in about twenty more LMXBs
(see van der Klis 2000, 2004, for a review). The kHz
QPOs in the power spectra of these systems cover the range of frequency
from some hundred Hz to more than one  kHz, and they often occur
in pairs in the persistent emission: the upper kHz QPO frequency
($\nt$, hereafter the upper-frequency) and the lower kHz QPO
frequency ($\no$, hereafter the lower-frequency).
  The kHz QPOs  were soon found to behave in a
rather regular way and the study of their phenomenology led to
the discovery of tight correlations between. their frequencies and
other observed characteristic frequencies (see, e.g.,
 Psaltis et al. 1998, 1999ab;  Stella et al.  1999; Belloni \et \,2002).
 Without any doubt,
{\em RXTE} has provided a probe into the accretion flow in the
non-Newtonian strong gravity regime where Einstein's  General
Relativity might be tested (van der Klis 2000, 2004).

The correlation between the upper-frequency and the lower-frequency
across different sources or for a particular  source, such as Sco X-1,
 can be roughly fitted  by a  power law function (see, e.g.,
 Psaltis et al. 1998, 1999a), but also by a
linear model (see Belloni et al. 2005).  The kHz QPO peak
separation $\dn\equiv\nt-\no$ between the upper-frequency and
lower-frequency   in a given source is generally inconsistent with
a NS  spin frequency. In some sources,
  $\dn$ is lower or higher than the NS spin frequency (when
  directly measured, see e.g. M\'endez \& van
der Klis 1999; Jonker, M\'endez, \& van der Klis 2002b) or than the
 nearly coherent oscillation frequency ($\nb$) observed
during type~I X-ray burst that is identified to be the stellar
spin frequency (Muno 2004; Strohmayer \& Bildsten 2003; Wijnands
et al 2003;  van der Klis 2004).
 The averaged
 peak separations  are found to be either
 close to (but not consistent with) the spin frequency $\ns$ or to its half $\ns$/2
 (see, e.g., Wijnands \et 2003;
 van der Klis 2004; Lamb \& Miller 2001).
 The above observations offer strong evidence   against the
 simple
beat-frequency model, in which the lower-frequency
is the beat  between the upper-frequency and the NS
 spin frequency $\ns$ (see e.g. Strohmayer et al.\ 1996; Zhang et al. 1997;
 Miller et al. 1998), i.e. $\no=\nt-\ns$.
 Nonetheless, with the discovery of  pairs of 30--450 Hz QPOs  from a few
black-hole candidates with  frequencies ratios 3:2 (see, e.g., van
der Klis 2004),   Abramowicz \et  (2003ab) reported that the ratios
of twin  kHz QPOs in  Sco X-1 tend to cumulate  around a value of
3:2, and they interpreted  it  as  an evidence of  a
 near $\sim$  3:2 resonance.
 This was further argued by  Abramowicz \et (2003ab) to be
 a promising link with the black-hole high-frequency QPOs (see e.g.
van der Klis 2004).
Moreover, the
production mechanisms of kHz QPOs are  still  open
issues: they have been identified with various characteristic
frequencies in the inner accretion flow (see e.g. Stella \& Vietri
1999;  Titarchuk et al. 1998;  Titarchuk \& Osherovich 2000;
Psaltis \& Norman 2000; Lamb \& Miller 2001; Zhang 2004).

 In this paper,  in order to check the  predictions
of the kHz QPO models,
 we analyze  the recently  published   kHz QPO data  by {\em RXTE\/},
   which have been used by several authors (see, e.g.,
   Belloni et al. 2005;  M\'endez \& van der Klis  1999, 2000;
   Psaltis et al. 1998, 1999ab;  van der Klis 2000,
 2004, and original references therein). Most of the  data
  are   provided by T. Belloni, M. M\'endez and D. Psaltis, and
  the others  are   extracted  from  the references listed
 in Table 1.
 Therefore, the  data we analyzed  here constitute a larger sample
than  that presented  by  Belloni  \et  (2005).
In section 2, we  critically discuss the
  twin  kHz QPO correlation.
   The conclusions and discussions are
     given  in the last section.

\section{The Correlations between the twin  kHz QPOs}

Figure~1 shows various correlations of the twin
kHz QPOs, such as  $\no$ vs. $\nt$,   $\dn$ vs. $\nt$   and
$\nt/\no$ vs. $\nt$, obtained using the simultaneously detected twin kHz
QPO data of LMXBs
listed in Table 1,
 We fitted  power-law relation,
  \begin{equation}
 \nu_1= a
  \left(\frac{\nu_2}{1000~\mbox{Hz}}\right)^{b}~\mbox{Hz}\;, \;\;\;
 \label{nu12}
 \end{equation}
   for the kHz QPO samples of  Atoll and Z  sources separately.
  A similar relation  was discussed
   by Psaltis et al. (1998), with a smaller set of QPO data points
 for Sco X-1.
  The normalization coefficient
 $a$ and the power-law index $b$ for various cases are listed in
 Table 2. We find that the  $\no$ vs. $\nt$ correlations for
the Atoll source 4U 1728-34 and  the Z
 source Sco X-1
  are somewhat distinct in their power-law indices, $\sim$1.5  and
 $\sim$1.9, respectively.
   Furthermore,  for different ranges  of $\nt$
  in  Z sources   (i.e. $\nt < $ 840  Hz and $\nt > $ 840  Hz),
 we obtain different normalization coefficients
  and power-law indices (i.e. 2.20 and 1.79). However, the $\chi^2$
  test in figure \ref{fig-chi}a seems not to favor  this broken
  power-law
  correlation because of the slightly larger $\chi^2$-values.
 Figure \ref{fig-chi}a shows  $\chi^2$  tests  that correspond to
the relation~(\ref{nu12}), where in many cases the minimum   $\chi^2$-values
 are  larger than  $\sim 3.0$  for the Z and Atoll samples,
 relatively too high values  consider them good fits.
 However, using only the data points for  Sco X-1 and
 4U 1728-34,
 the minimum $\chi^2$-values  for
relation~(\ref{nu12}) become $\simeq 0.6$ and  $\simeq 0.9$
respectively, less than the
corresponding value  $\sim 2$ for Sco X-1 previously obtained by
Psaltis et al (1998). Therefore, for the individual Z or Atoll
sources Sco X-1 and 4U 1728-34, we obtain
power-law correlations between
the kHz QPOs
 with different normalization coefficients  power-law indices
 between the two sources.
%
The choice of a power-law  relation to describe the kHz QPO data
is arbitrary. Other
functional forms can also fit the data equally well, as
the linear correlation adopted by Belloni et al. (2005) with
a smaller sample of QPO data. If relation
Eq.(\ref{nu12}) is valid,
 it implies  a non-monotonic change of the peak
separation, with a maximum at an upper kHz QPO frequency of $\sim
700$~Hz, as shown in Figure 1b.
%


\begin{figure}
   \hspace{-8.mm}
   \begin{center}
\includegraphics[width=7.5cm]{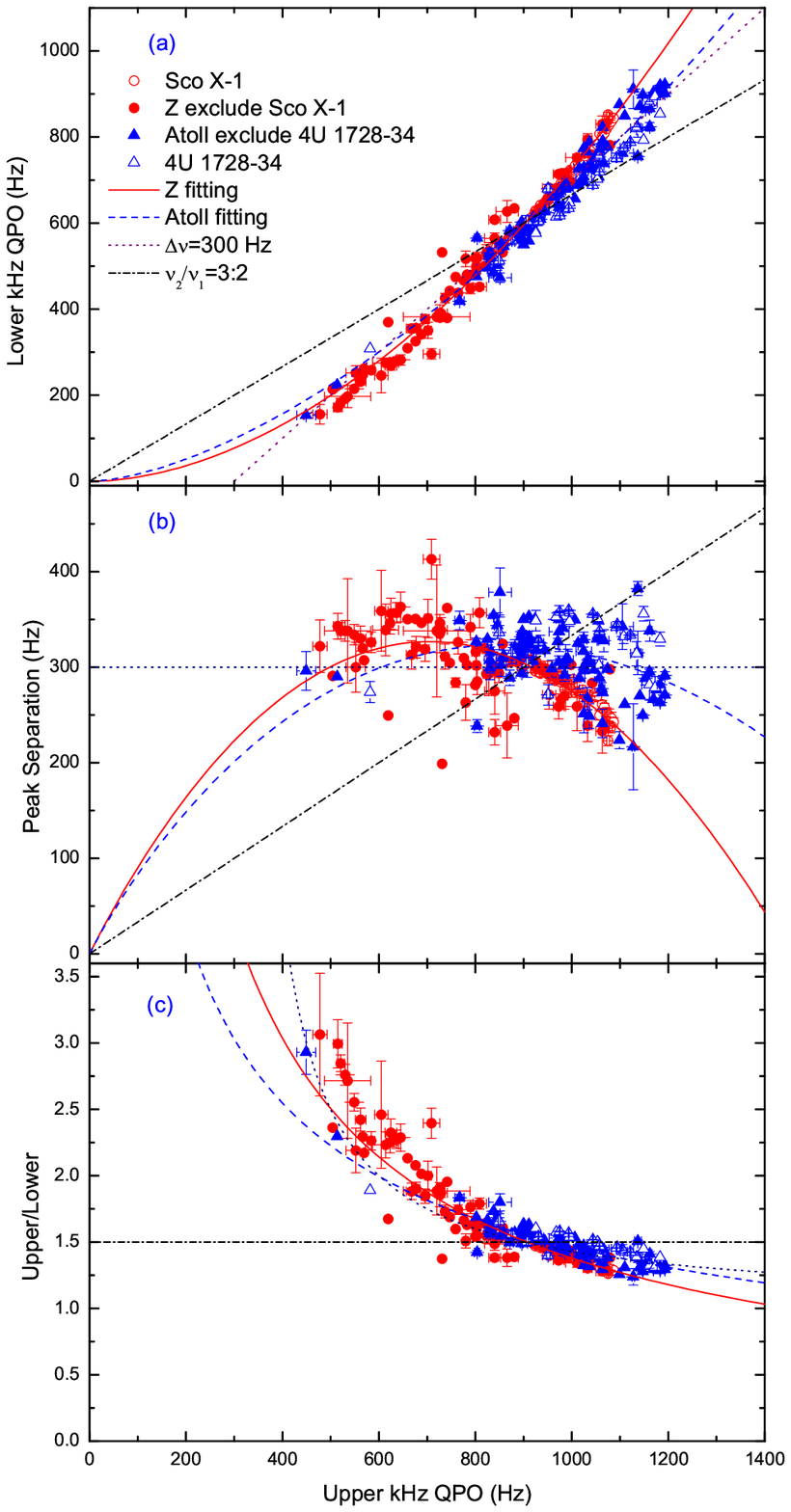}
\caption[fig1] { Plots of (a) $\no$ vs. $\nt$,  (b) $\dn$ vs.
$\nt$ and  (c) $\nt/\no$ vs. $\nt$.  The Z [Atoll] fitting line
represents the fitted correlation between the pair kHz QPO
frequencies for the Z [Atoll] sources as $\no=(724.99\pm2.52
\;{\rm Hz}) (\nt/1000 {\rm Hz})^{1.86\pm0.03}$
[$\no=(683.48\pm3.01 \;{\rm Hz}) (\nt/1000 {\rm
Hz})^{1.61\pm0.04}$].  } \label{fig-12}
   \end{center}
\end{figure}

\begin{figure}
   \hspace{-8.mm}
   \begin{center}
\includegraphics[width=7.5cm]{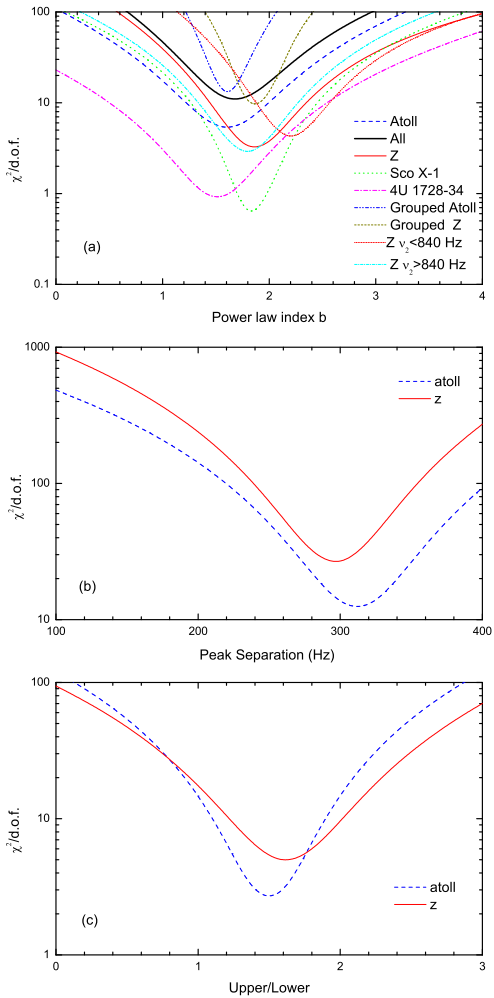}
\caption { Chi-square tests for  (a)  the hypothesis of a
power-law correlation, (b) a constant peak separation  and (c)
a constant ratio  between the twin  kHz QPOs of   LMXBs listed
in Table 1. }
 \label{fig-chi}
   \end{center}
\end{figure}

\subsection{Testing the constant twin peak separation}

 In fact,  it is not just for Sco X-1 (van der Klis et al 1997;
M\'endez \& van der Klis 2000) that the peak separation is known to
be not constant but also for other Z sources, e.g.,   GX 17+2 (Homan
et al. 2002),  GX 340+0 (Jonker et al. 2000) and GX 5-1 (Jonker et
al. 2002a),   and for several Atoll sources, e.g., 4U~1728--34
(Migliari, van der Klis,  \& Fender
 2003; M\'endez \& van der Klis 1999), with $\dn$  always
significantly lower than the burst oscillation frequency $\nb$,
4U~1636--53 (Jonker, M\'endez, \& van der Klis 2002b; M\'endez,
van der Klis, \& van Paradijs 1998b) with  $\dn$ varying between being
lower and higher than $\nb/2$, 4U 1608-52 (M\'endez et al. 1998c)
 and   4U 1735-44 (Ford et al. 1998), etc.

 In order to test the hypothesis of a constant peak separation
 between the twin kHz QPOs, we inspect our sample of  simultaneously
detected twin kHz QPOs over a wide range of  frequencies separately for
Atoll sources (110 pairs) and Z sources (158 pairs).
%
 In figure~\ref{fig-12}b,  we show that the peak separation in
individual Atoll or Z sources, as for instance Sco~X-1 and 4U
1728-34,
 decreases   systematically with increasing upper frequency if $\nt$
is larger  than $\sim$700 Hz, a fact that has been reported in the
literatures before (see, e.g. van der Klis 2000, 2004 for
recent reviews).
%
%
%
%
Figure \ref{fig-chi}b  shows the results of a $\chi^2$ test against a
constant peak separation of twin  kHz QPOs in Atoll and Z
  sources.   The
resulting minimum $\chi^2$-values are very high, and we
therefore conclude that there is not
 a constant peak separation
 either for Z sources or Atoll sources.
%
%
 Nevertheless, this also confirms the previously known result that
 the Sco~X-1 data are inconsistent
with a constant peak separation (see also van der Klis et al.\
1997; Psaltis et al. 1998; M\'endez \& van der Klis  2000).
%

\subsection{Testing a preferred 3:2 ratio in the twin kHz QPOs}

From Figure \ref{fig-12}b, one can see that a constant ratio
relation $\nt=(3/2)\no$, shown as a dash-dotted line, is not
consistent with the observed data.
Moreover,
Figure \ref{fig-12}c shows that the frequency ratio  systematically decreases
from 3.2 to 1.2 with increasing  the kHz QPO frequency.
  As a further investigation,
 we also performed a $\chi^2$ test against a constant ratio,
 shown in figure \ref{fig-chi}c.
    The obtained  $\chi^2$-values   are too  high to
    confirm a 3:2 peak ratio for the all kHz QPO  data.
    In addition,
 the incompatible  3:2 ratio peak distribution
  has been also studied  by Belloni et al. (2005) for many sources:
  they showed that  the distribution of QPO frequencies in {Sco X-1},
 {4U 1608--52},  {4U 1636--53},  {4U 1728--34}, and
 {4U 1820--30} is multi-peaked, with the peaks occurring at
 the different $\nu_2 / \nu_1$ ratios,
 not all ratios appearing in all sources.

%

\subsection{Testing other kHz QPO models}
\begin{figure}
   \begin{center}
\includegraphics[width=7.5cm]{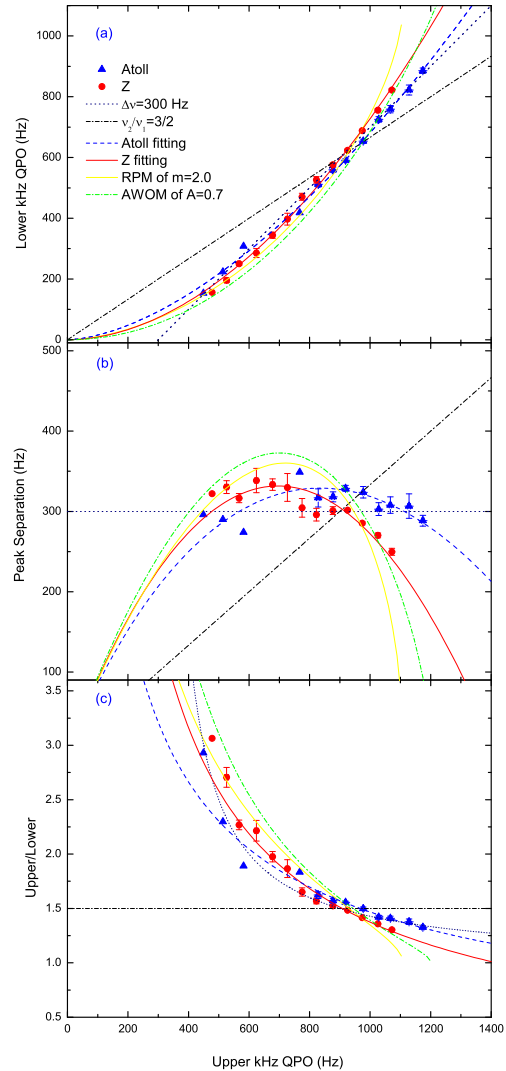}
\caption[fig1] { The panels are the same as for
Figure \ref{fig-12}. For reason of
 clarity, we divided  all
kHz QPO data into  bins of 50 Hz interval in $\nt$, and then
average the corresponding kHz QPO  data in each bin. The Z
(Atoll) fitting  line represents the fitted correlation between
the twin kHz QPOs for grouped Z (Atoll) samples as
$\no=(727.88\pm6.63 \;{\rm Hz})(\nt/1000 {\rm Hz})^{1.91\pm0.05}$
($\no=[681.42\pm5.39 \;{\rm Hz}][\nt/1000 {\rm
Hz}]^{1.65\pm0.05}$). AWOM (RPM) is the theoretical curve of the
model by Zhang (2004) with the averaged stellar mass density
parameter A=0.7 (by Stella \& Vietri [1999] with the mass
parameter m=2.0 \ms).
  }\label{fig-50}
   \end{center}
\end{figure}

In order to account for  a variable peak separation
   of twin kHz QPOs,   some viable models have been proposed.
The variable peak separation was predicted
 in the relativistic precession model (Stella \& Vietri 1999)
and the Alfv\'en wave oscillation model (Zhang 2004), where for both models the
upper-frequency was ascribed to the Keplerian orbital frequency,
while the lower-frequency was  ascribed to the
periastron precession frequency and the Alfv\'en wave oscillation
frequency, respectively.
In Figure \ref{fig-12}, the QPO data points are scattered, which makes it
difficult to estimate the kHz QPO relations by reading
the figure directly.  In order to compare clearly the models
with the trends of twin kHz QPOs, we divided all data points into
50 Hz bins, and then averaged the quantities
in every bin. We plot these group-averaged values in
Figure \ref{fig-50}, with the same panels as
Figure \ref{fig-12}.
We point out that neither the relativistic precession model nor
the Alfv\'en oscillation model can explain the distinctions of
kHz QPOs for both the Atoll  and  Z sources, even though
both models are in good agreement with the observed kHz QPO data,
as shown in Figure \ref{fig-50}, once the model parameters
are tuned.

The motivation for  performing  a 50 Hz binning  to discuss
the twin kHz QPO correlations was that
 we want to show the averaged effects.
 The dispersions of the averaged data points  are
  bigger than their averaged  error  bars (Figure \ref{fig-50}),
  so the minimum  $\chi^2$-value
    corresponding to each group  is usually much larger than 1.0.
  Therefore, we remark that
  their error bars underestimate the true
uncertainties, and  the calculations show that  the
 minimum   $\chi^2$-values   for the fitted lines
 of the grouped samples, as shown in the caption of Figure \ref{fig-50},
  are  also too high for the fits to be acceptable.

\section{ CONCLUSIONS AND DISCUSSIONS}

 We have analyzed an updated  sample of frequencies of the
simultaneously detected twin kHz QPOs in LMXBs.
Our main conclusions  are the following.
(1)
 The power-law correlations were analyzed by means  of $\chi^2$ tests:
  the  sources 4U 1728-34 and Sco X-1
  are found to yield good power-law fits, with minimum
$\chi^2$-values  lower than 1.
  The   power-law indices are
   $b\simeq1.5$ for the   Atoll  source 4U 1728-34
  and  $b\simeq1.9$  for the Z  source  Sco X-1.
 A similar power-law index was previously obtained
  with a slightly high
    minimum $\chi^2$-value $\sim 2$ by Psaltis et al. (1998).
 As it is known,  Atoll  and  Z sources show distinct
properties in their spectra and luminosity (Hasinger \& van der
Klis 1989; van der Klis 2000),
 and we do not yet know what properties cause
 the  differences in their  power-law indices.
 Nevertheless, if the power-law relations with different
indices for Atoll and Z sources are confirmed by
future detections, then
 any kHz QPO models discarding the distinctions of the Atoll
and Z sources will confront severe arguments.
%
(2) Clearly, obeying such a power-law relation would contradict
 the constant peak separation and constant (3:2) peak
ratio between kHz QPOs,  and in fact the plotted curves
in Figure \ref{fig-12} and Figure \ref{fig-50}
  are incompatible with these constant relations.
 These  conclusions have been previously inferred with smaller samples of
 kHz QPO data (see, e.g. Psaltis et
al. 1998; Psaltis et al. 1999ab; Belloni
et al. 2005), but contrary to
 the suggestions by the simple beat model and   any   model
  that predict
   $\dn$=constant  and   $\nt =(3/2)\no$, respectively.
 In addition, based on  the updated kHz QPO data  of LMXBs
  we find  that there is no extremely
 sharp concentration at  a 3:2 peak ratio as indicated
by the $\chi^2$ test in Figure 2c; the
   ratios are broadly distributed  from $\sim 1.2$
    to $\sim 3.2$ over a  frequency
   range of some hundred Hz, as shown in Figure \ref{fig-12}c.
   Therefore,   the  non-linear resonance model
  (see e.g. Abramowicz et al. 2003b; Rebusco 2004) can be
   consistent with
  this  distribution of peak ratios. Nevertheless, it is shown
in Figure 1c that the value of $\nt/\no$ decreases systematically
 with increasing QPO frequency.
(3) In a rough approximation, the kHz QPO frequency correlation
 seems to be consistent with the predictions by  the model based on the
basic general relativistic frequencies around a compact object
(Stella \& Vietri 1999) and those by the model  based on the Alfv\'en wave
oscillation (Zhang 2004), with properly selected  parameters.
Both models predicted a peak separation decreasing with
increasing QPO frequency, but increasing when the upper
frequency is low, for instance less than  $\sim$700 Hz.
Therefore,
more kHz QPO detections are needed to confirm the predictions
of the models.
%
In conclusion,  if future data still support the conclusions
 obtained in the paper,
 they will pose new constraints on models for  explaining kHz QPOs.

\section{acknowledgements}
We are grateful for  T. Belloni,  M. M\'endez and D. Psaltis     for
providing the QPO data,
and many helpful  discussions with T.P. Li, M. Abramowicz, T.
Belloni,  Klu\'zniak, W., P. Rebusco and J. Petri  are highly
appreciated.
%
This research has been supported by the innovative project of CAS
of China.
The authors express   the  sincere thanks to the critical comments
 from the anonymous referee  that greatly improved the
 quality of the paper.
%


\newpage

\begin{figure*}
\hskip -2.5cm
\centerline{\psfig{file=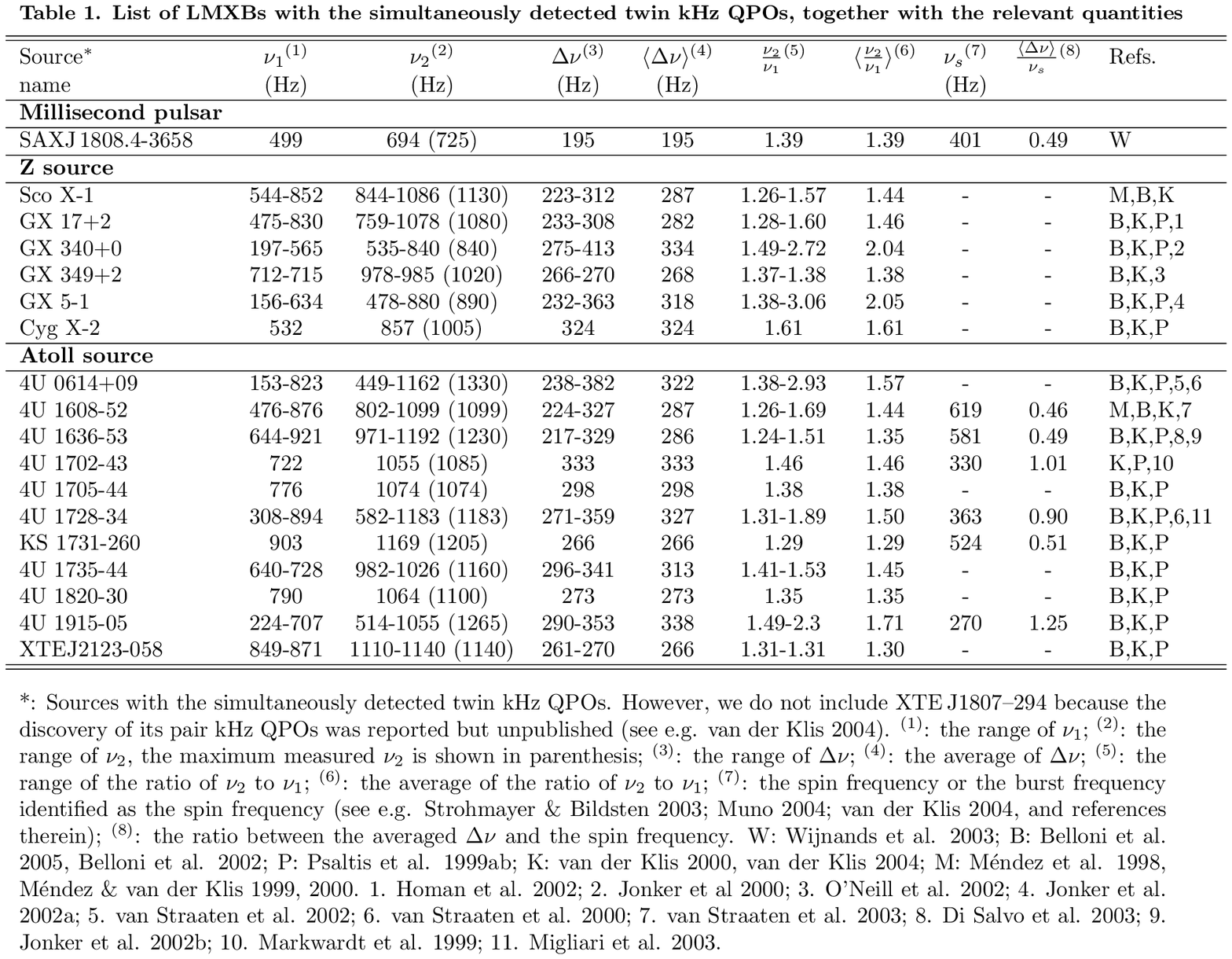,angle=0,height=24cm,width=440pt}}
\end{figure*}

\begin{tabular}{c}
{\small \bf  Table 2. The   twin kHz QPO correlation
$\no=a(\nt/1000 {\rm Hz})^{b}$}
\end{tabular}

\begin{tabular}{lcc}
\hline \hline
Sources$^{*}$   &   $a$ (Hz) &  $ b$ \\
\hline 
{\rm  All Z samples}     & 724.99$\pm$2.52 & 1.86$\pm$0.03\\
 {\rm  Z ($\nt<840 \;{\rm Hz}$)}   & 812.09$\pm$2.59 & 2.20$\pm$0.10\\
 {\rm  Z ($\nt>840 \;{\rm Hz}$)}   & 722.72$\pm$1.45 & 1.79$\pm$0.03\\
 {\rm    Sco X-1}     & 721.95$\pm$0.69 & 1.85$\pm$0.01\\
{\rm All Atoll samples }     & 683.48$\pm$3.01 & 1.61$\pm$0.04\\
{\rm   4U 1728-34}     & 667.86$\pm$5.59 & 1.51$\pm$0.07\\
\hline
All Z and Atoll samples  & 699.13$\pm$2.23 & 1.68$\pm$0.02\\
\hline
\end{tabular}

\begin{tabular} {c}
 $^*$: Data are taken from the references listed in Table 1.
\end{tabular}

\end{document}